\documentclass[letter]{aa} 

\usepackage{graphicx}
\usepackage{txfonts}
\usepackage{xcolor}

\makeatletter
\renewcommand*\aa@pageof{, page \thepage{} of \pageref*{LastPage}}
\makeatother

\begin{document}

   \title{A self-synthesized origin of heavy metals in hot subdwarf stars?}


   \author{Battich T.
          \inst{1}
          \and
          Miller Bertolami M. M.
          \inst{2,}\inst{3}
          \and
          Serenelli A. M.
          \inst{4,}\inst{5}
          \and       
          Justham S.\inst{1,}\inst{6}
          \and       
          Weiss A.\inst{1}
          }

   \institute{Max-Planck-Institut f\"ur Astrophysik, Karl-Schwarzschild Strasse 1, D-85748, Garching, Germany\\
              \email{tiara@mpa-garching.mpg.de}
         \and
          Instituto de Astrofísica de La Plata, Consejo Nacional de Investigaciones Científicas y Técnicas Avenida Centenario (Paseo del Bosque) S/N, B1900FWA La Plata.
         \and
          Facultad de Ciencias Astronómicas y Geofísicas, Universidad Nacional de La Plata Avenida Centenario (Paseo del Bosque) S/N, B1900FWA La Plata, Argentina
        \and Institute of Space Sciences (ICE, CSIC), Carrer de Can Magrans S/N, E-08193, Cerdanyola del Valles, Spain
          \and Institut d’Estudis Espacials de Catalunya (IEEC), Carrer Gran Capita 2, E-08034, Barcelona, Spain
          \and Anton Pannekoek Institute for Astronomy, University of Amsterdam, 1090 GE Amsterdam, The Netherlands
             }

   \date{Accepted for publication in Astronomy and Astrophysics Letters}

 
  \abstract
   {Some He-rich hot subdwarf stars present high abundances of trans-iron elements, such as Sr, Y, Zr and Pb. Diffusion processes are important in hot subdwarf stars, and it is usually thought that the high abundances of heavy elements in these peculiar stars are due to the action of radiative levitation. However, during the formation of He-rich hot subdwarf stars, hydrogen can be ingested into the convective zone driven by the He-core flash. It is known that episodes in which protons are being ingested into He-burning convective zones can lead to neutron-capture processes and the formation of heavy elements.} 
   {In this work we aim to explore for the first time if neutron-capture processes can occur in late He-core flashes happening in the cores of the progenitors of He-rich hot subdwarfs. 
   We aim to explore the possibility of a self-synthesized origin of the heavy elements observed in some He-rich hot subdwarf stars. }
   {We compute a detailed evolutionary model of a stripped red-giant star with a stellar evolution code with a nuclear network comprising 32 isotopes. Then we post-process the stellar models in the phase of helium and hydrogen burning with a post-processing nucleosynthesis code with a nuclear network of 1190 species that allows us to follow the neutron-capture processes in detail.}
   {We find the occurrence of neutron-capture processes in our model, with neutron densities reaching a value of $\sim5\times10^{12}\,{\rm cm}^{-3}$. 
   We find that the trans-iron elements are enhanced in the surface by 1 to 2 dex as compared to initial compositions.
   Moreover, the relative abundance pattern $[{\rm X}_i/\rm{Fe}]$ produced by neutron-capture processes closely resembles those observed in some He-rich hot subdwarf stars, hinting at a possible self-synthesized origin of the heavy elements in these stars.}
   {We conclude that intermediate neutron-capture processes can occur during a proton ingestion event in the He-core flash of stripped red-giant stars. This mechanism offers a  natural channel to produce the heavy elements observed in some of the He-rich hot subdwarf stars.}

   \keywords{ nuclear reactions, nucleosynthesis, abundances --
            subdwarfs     --
             stars: abundances -- 
             stars: chemically peculiar --
             stars: evolution --
             stars: interiors
               }

   \maketitle
%

\section{Introduction}

The majority of the hot subdwarf stars (sdOBs) are low-mass He-core burning stars with a thin, radiative, hydrogen (H)-rich envelope  \citep{2016PASP..128h2001H}. Their atmospheres are composed mostly of pure H due to the action 
of diffusion processes, in particular gravitational settling. However, a fraction of them are helium (He) rich \citep{2017A&A...600A..50G}. He-rich hot subdwarfs (He-sdOB) are believed to be formed via the merger of two He-core white dwarfs (WDs) \citep{Webbink1984WDmergers, 2012MNRAS.419..452Z, 2016MNRAS.463.2756H, 2018MNRAS.476.5303S}, a He-core WD with a low-mass CO-core WD \citep{Justham+2011He-sdO, 2022MNRAS.511L..60M}, or in interactive
binaries where a red giant lost most of its H-rich envelope in a mass-transfer episode and/or after a common-envelope episode \citep{2003MNRAS.341..669H, 2020A&A...642A..97K}. Moreover, the formation of single hot subdwarf stars can happen after the engulfment of a substellar companion thanks to the positive binding energy of the stellar envelope near the tip of the red giant branch \citep{2010AIPC.1314...85H, 2020MNRAS.496..612H}. In addition to these scenarios, a fully single stellar evolution scenario for these stars is also possible in stars with low initial masses and metallicities but high initial He contents  \citep[$M_i\sim 0.7M_\odot$, $Z\sim 0.001$, and $Y\sim 0.4$,][]{2017A&A...597A..67A}. Regardless of the specific evolutionary scenario, in the cases in which the star has lost almost all of its H-rich envelope near the tip of the red giant, the He-flash driven convective zone reaches the H-rich material and burns most of it making the star He-rich at the surface \citep{2008A&A...491..253M}. 
A small group of hot subdwarf stars present a mixture of H and He in their atmospheres, and are referred to as intermediate He-rich hot subdwarfs (iHe-sdOBs, see \citealt{2012MNRAS.423.3031N}). It has been suggested that these stars might be in a transition phase of their evolution in which they are evolving from the He-flash towards the horizontal branch, and their atmospheres are transitioning from a He-dominated composition into a H-dominated one due to diffusion processes \citep{2010MNRAS.409..582N}. 

More than a decade ago, \cite{2011MNRAS.412..363N} detected high abundances of heavy elements in LSIV-14$^{\circ}$116, an iHe-sdOB. In particular, they reported abundances in number fraction of a thousand times the solar value of zirconium (Zr), yttrium (Y) and strontium (Sr), and a hundred times of germanium (Gr). \cite{2011MNRAS.412..363N} suggested that the high abundance of these elements might be due to the action of radiative levitation pushing the elements towards the line-forming region of the star. They also discussed that the presence of a strong magnetic field could both explain the abundances anomalies as well as the light variability that is observed in this star. However, \cite{2015A&A...576A..65R} ruled out the latter placing an upper limit of 300G for the magnetic field of LSIV-14$^{\circ}$116. After the discovery of the peculiar abundances in LSIV-14$^{\circ}$116, the number of known iHe-sdOB stars with high abundances of heavy elements has increased to eleven, namely, HE 2359-2844, HE 1256-2738 \citep{2013MNRAS.434.1920N}, UVO 0825+15 \citep{2017MNRAS.465.3101J}, EC 22536-5304 \citep{2019MNRAS.489.1481J,2021A&A...653A.120D}, Feige 46 \citep{2019A&A...629A.148L}, HZ 44, HD 127493 \citep{2019A&A...630A.130D}, PG 1559+048, FBS 1749+373 \citep{2020MNRAS.491..874N}, PHL 417 \citep{2020MNRAS.499.3738O}, and LSIV-14$^{\circ}$116 itself, whose abundances have been revised with non-local thermodynamic equilibrium (NLTE) atmospheric models by \cite{2020A&A...643A..22D}, confirming the high heavy-metal enrichment. 

Radiative levitation plays a key role in setting the photospheric abundances of hot subdwarf stars in general (e.g., \citealt{2011A&A...529A..60M}). Therefore, this process almost surely plays a role in determining the metal abundances of iHe-sdOBs. If radiative levitation is at work, as noted by \cite{2011A&A...529A..60M}, two key questions need to be answered: whether evolutionary timescales are long enough for radiative levitation to enhance their abundances significantly, and whether there is enough material in neighboring layers for diffusion to act efficiently. Both questions are connected as different initial abundances would affect the time required by diffusion to reach the abundances observed in individual iHe-sdOB stars. These stars have kinematics consistent with Halo objects (see, e.g., \citealt{2015A&A...576A..65R,2019A&A...629A.148L,2019A&A...630A.130D}), pointing to a low metallicity origin which would imply that they have been born with lower abundances of trans-iron elements than the field sdOB stars studied by \cite{2011A&A...529A..60M}. Therefore, the question of whether enough heavy elements are present in the stars for radiative levitation to act efficiently is particularly relevant for the heavy-metal iHe-sdOBs. Alternatively, however, the Halo-kinematics of these stars might be explained if they are the surviving companions of Type Ia supernovae \citep[][and references therein]{2021MNRAS.507.4603M}.

A self-synthesized origin of the heavy elements in these stars have been ruled out mainly due to the lack of evidence of them having evolved through the thermally pulsing phase of asymptotic giant branch (AGB) stars \citep{2011MNRAS.412..363N,2013MNRAS.434.1920N}. AGB stars are known to produce heavy elements by means of the slow neutron-capture process during their thermally pulsating phase (see, e.g., \citealt{2014PASA...31...30K} for a review and references therein). Neutron-capture processes are responsible for the formation of most of the elements heavier than iron. Two main types of neutron-capture processes can be identified: the slow ($s$-) and rapid ($r$-) neutron-capture processes, which differ mainly in the neutron density of the environments in which they take place ($N_n \sim 10^7 - 10^{11}\,$cm$^{-3}$ for the $s$-process and $N_n\gtrsim 10^{20}\,$cm$^{-3}$ for the $r$-process, \citealt{1957RvMP...29..547B}) and therefore in the abundance patterns of heavy elements they produce. \citet{1977ApJ...212..149C} have proposed that neutron-capture nucleosynthesis might happen inside stars at intermediate values of neutron densities, $N_n \sim 10^{12}-10^{16}\,$cm$^{-3}$, a process named intermediate ($i$-) neutron-capture process. The $i$-process has been proposed to explain the abundance pattern of some carbon-enhanced metal-poor stars which cannot be explained by a combination of $r$- and $s$-process enrichment (e.g. \citealt{2012AIPC.1484..111L,2015arXiv150505500D}) and the abundance pattern observed in Sakurai's Object, a post-AGB star \citep{2011ApJ...727...89H}. The proposed astrophysical sites for the $i$-process include the first thermal pulses of very low-metallicity AGB stars \citep{2010A&A...522L...6C,2021A&A...648A.119C, 2021A&A...654A.129G, 2022A&A...667A.155C}, rapidly-accreting WDs \citep{2017ApJ...834L..10D,2019MNRAS.488.4258D, 2021MNRAS.503.3913D, 2018ApJ...854..105C, 2021MNRAS.504..744S}, the main He flash of ultra low-metallicity stars in the red giant branch \citep{2010MNRAS.405..177S, 2010A&A...522L...6C, 2013A&A...559A...4C}, low-metallicity massive stars \citep{2018ApJ...865..120B,2018MNRAS.474L..37C,2021MNRAS.500.2685C} and very late thermal pulses in post-AGB stars \citep{2011ApJ...727...89H, 2018JPhG...45e5203D}. All of these sites have in common that the $i$-process nucleosynthesis happens when protons are being ingested into a convective He-burning zone. In this environment, $^{13}$C is created, and a flux of neutrons can be produced via the $^{13}{\rm C}(\alpha, n)^{16}\rm{O}$ reaction. He-rich hot subdwarf stars can be formed when the He-core flash occurs in a stripped red-giant star. As mentioned above, in this case, the convective zone driven by the He flash can reach the H-rich layers and ingest protons into the He-burning zone, potentially creating a flux of neutrons. In the double merger scenario, also, He is ignited in off-centered flashes in the post-merger remnant before the quiescent core-He burning phase \citep{2012MNRAS.419..452Z,2018MNRAS.476.5303S}. As pointed out previously by \cite{2012MNRAS.419..452Z}, the convective zone of these He-flashes might also reach the proton-rich surface, depending on the mass of H that survives the merger, which is an uncertain quantity \cite[see, e.g., ][]{2016MNRAS.463.2756H,2018MNRAS.476.5303S}. Up to now, however, the possibility of $i$-process nucleosynthesis and heavy elements production in this scenario suitable for the formation of He-rich sdOBs has not been studied. Therefore, a self-synthesized origin of the heavy elements observed in the iHe-sdOB stars is a realistic possibility and needs to be explored. 

In this letter, we compute detailed nucleosynthesis calculations in a model of a late He-core flash and show that heavy elements can be produced due to neutron-capture processes, which will appear later at the surface of the star. We discuss this result in the light of the abundance pattern observed in hot subdwarf stars. 
The paper is organized as follows. In section \ref{sec:2} we describe the stellar model and method used for the nucleosynthesis calculation. In section \ref{sec:3} we present our results and discuss them, and in section \ref{sec:4} we summarize our conclusions.


\section{Method and model}\label{sec:2}

Most of the hot subdwarf stars have masses lower than $1\,M_{\odot}$ (see, e.g., \citealt{2016PASP..128h2001H} and references therein). For modeling a hot subdwarf star, we have calculated the evolution of a $1\,M_{\odot}$ stellar model from the main sequence to the red giant branch (RGB) tip. We have enhanced the mass loss close to the tip of the RGB in order to obtain the model of a stripped red-giant star. 
As mentioned in the introduction, the removal of the envelope can happen through winds in low-mass He-enriched single stars \citep{2005ApJ...621L..57L, 2017A&A...597A..67A}, by mass transfer in interactive binary systems, or  by the engulfment of substellar companions thanks to the low binding energy of the envelopes of RGB stars \citep{2010AIPC.1314...85H, 2020MNRAS.496..612H}.
 The final mass of our model is $0.4846\,M_{\odot}$ and a H-rich envelope mass of $6.09\times10^{-4}\,M_{\odot}$. Due to the action of diffusion processes in the atmosphere of hot subdwarf stars, it is not possible to determine their initial metallicity based on the surface abundances. However, as mentioned in the introduction, the iHe-sdOB stars enriched in heavy elements have kinematics compatible with Halo objects, hinting at a low metallicity origin. Moreover, \cite{2021A&A...653A.120D} reported that EC 22536-5304 is in a binary system, with a subdwarf F-type (sdF) star, for which they obtain a metallicity of $[{\rm Fe}/{\rm H}] = -1.95\pm 0.04$, with an $\alpha$-enhanced composition of $[{\rm \alpha}/{\rm Fe}]=0.4\pm 0.04$. Motivated by this finding, we have chosen for our models an initial metallicity of $Z = 0.0004$ and $[{\rm O}/{\rm Fe}]=0.4$. The stellar evolution calculations have been performed with the stellar evolution code \texttt{LPCODE} \citep{2005A&A...435..631A, 2016A&A...588A..25M}. The version of \texttt{LPCODE} used in this work has a nuclear network that comprises 32 species and 96 reactions (see Appendix A of \citealt{2020A&A...635A.164S}). For this work, the reaction rates are taken from the NACREII  compilation \citep{2013NuPhA.918...61X}.  

To study neutron-capture processes, we have recalculated the chemical evolution in a post-processing approach, with a reaction network that comprises 1190 nuclear species in order to be able to follow the nucleosynthesis when neutrons densities are as high as $N_n=10^{17}\,$cm$^{-3}$. These calculations have been performed with the nucleosynthesis post-processing code \texttt{ANT}: Astrophysical Nucleosynthesis Tool. \texttt{ANT} has been specifically developed during this work. The reaction rates for weak interactions (electronic captures and $\beta$ decays) are mainly taken from \cite{1987ADNDT..36..375T}, while all the other reaction rates are taken from the JINA Reaclib database \citep{2010ApJS..189..240C}. In \texttt{ANT} we include the screening corrections following the works of \cite{1973ApJ...181..439D} and \cite{1973ApJ...181..457G}. Convective mixing is calculated following the scheme of \cite{2001ApJ...554.1159C}. For more details of the code and the numerical approach see Appendix \ref{appx:1}. 


\section{Results and discussion}\label{sec:3}
\begin{figure}
	\includegraphics[width=1\columnwidth]{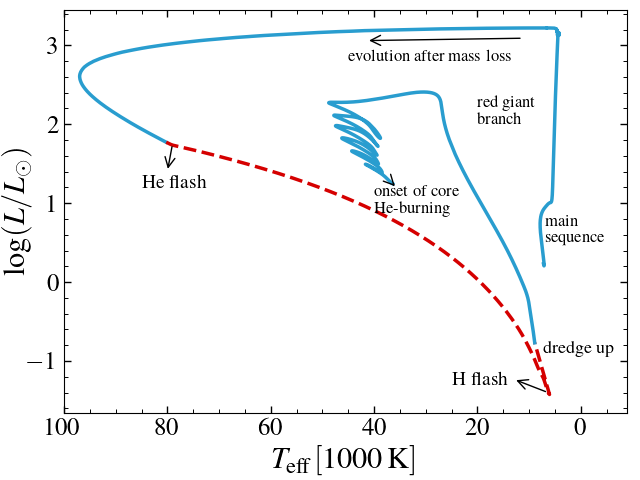}
    \caption{Hertzsprung-Russell diagram of the evolution of our model from the zero-age main sequence until the beginning of the quiescent core He-burning phase. The red dashed line corresponds to the phase of evolution shown in Fig. \ref{fig:kipp}.}
    \label{fig:HR}
\end{figure}
\begin{figure}
	\includegraphics[width=1\columnwidth]{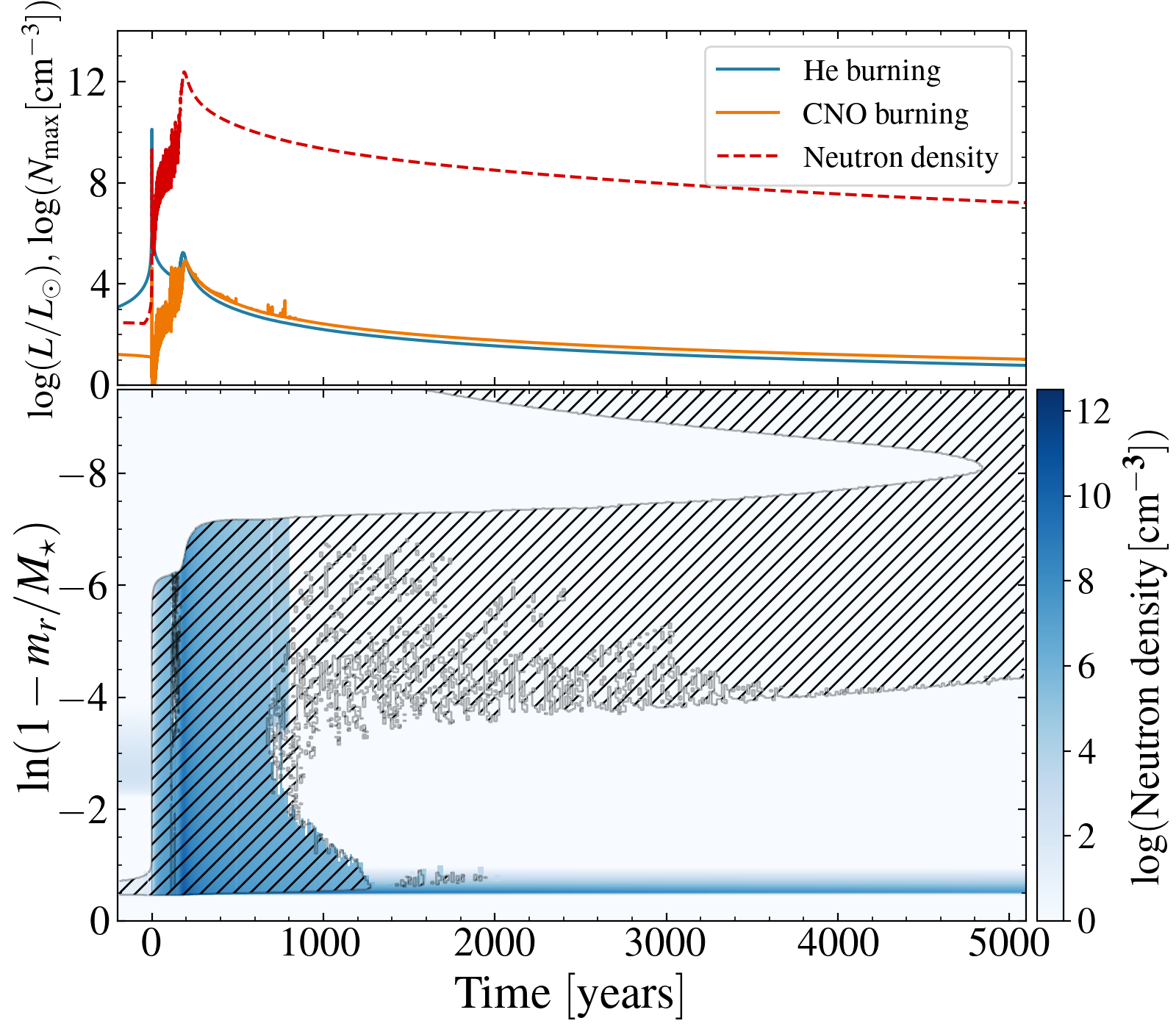}
    \caption{Evolution of different quantities of our model. Lower panel: Kippenhahn diagram showing the convective zones dashed in black, and the neutron density in ${\rm cm}^{-3}$ color-coded in shades of blue. The x-axis shows the time in years measured since the maximum release of energy of the He flash. The y-axis shows the outer mass in logarithmic scale, $\ln(1-m_r/M_{\star})$, being $m_r$ the mass coordinate and $M_{\star} = 0.4846\,M_{\odot}$ the mass of the star after the stripping. Upper panel: evolution of the luminosity due to He- (blue line) H-burning (orange line) respectively and of the maximum value of the neutron density (red dashed line).} 
    \label{fig:kipp}
\end{figure}
\begin{figure}
	\includegraphics[width=1\columnwidth]{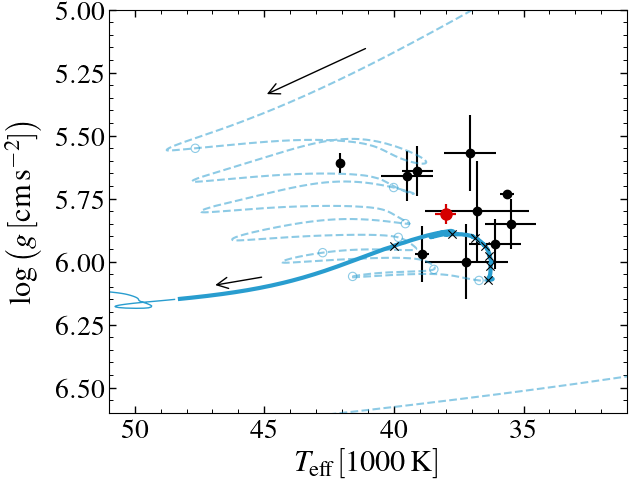}
    \caption{Kiel diagram showing the heavy-metal hot subdwarfs (black dots, \citealt{2013MNRAS.434.1920N,2020MNRAS.491..874N,2017MNRAS.465.3101J,2020MNRAS.499.3738O,2019A&A...630A.130D,2020A&A...643A..22D}) and our model. The dashed light line is the evolution before the star settles onto the horizontal branch. Light blue circles are placed here every $200\,$Kyr. The full thicker line is the evolution during the quiescent He-core burning phase. Black crosses are placed every $20\,$Myr in this part of the track. The red dot corresponds to the location of EC~22536-5304 \citep{2021A&A...653A.120D}.}
    \label{fig:kiel}
\end{figure}

\begin{figure*}
	\includegraphics[width=2\columnwidth]{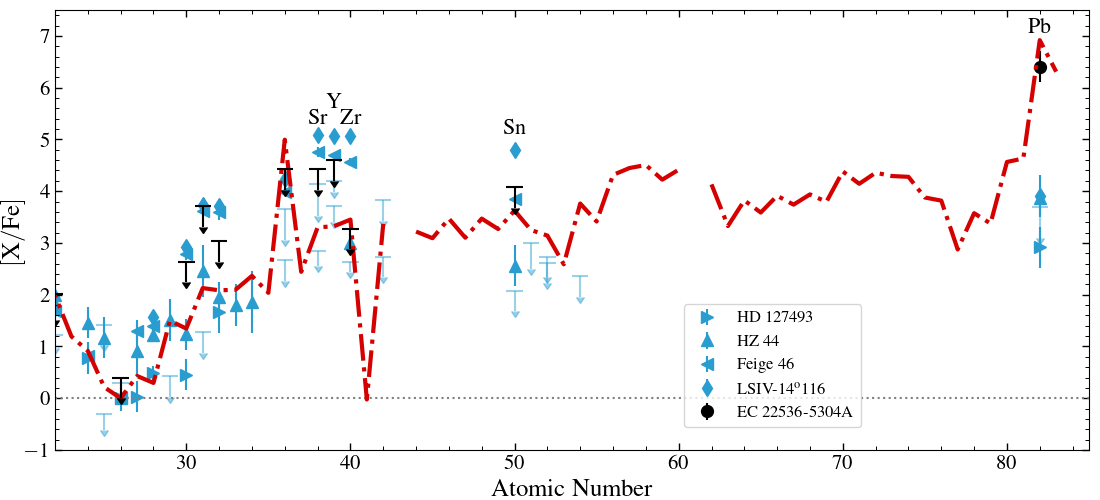}
    \caption{Abundances in number fraction respect to iron (and to the solar abundances). The prediction of our model is shown by the red line, while the points are the observations for 5 of the iHe-sdOB taken from \cite{2019A&A...630A.130D,2020A&A...643A..22D,2021A&A...653A.120D}. Different symbols correspond to different stars as indicated in the plot. The black symbols correspond to the abundances of EC 22536-5304. Upper limits are marked by horizontal lines with arrows pointing down. Solar values are taken from \cite{2009ARA&A..47..481A}.}
    \label{fig:abFe}
\end{figure*}

\begin{figure}
	\includegraphics[width=1\columnwidth]{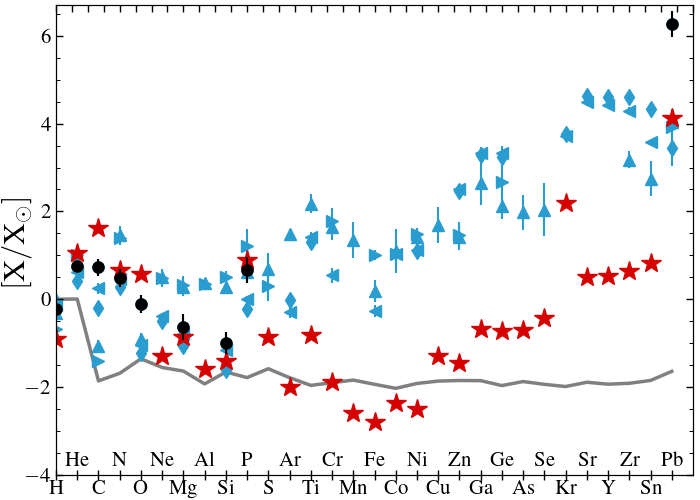}
    \caption{Abundances in number fraction respect to solar for the elements observed in the iHe-sdOB stars. The gray line corresponds to the initial abundances in our model. The predicted surface abundances after the dredge up are shown with red star symbols. Observed abundances are shown with the same symbols as in Fig. \ref{fig:abFe}. Solar values are taken from \cite{2009ARA&A..47..481A}.}
    \label{fig:absun}
\end{figure}

In Fig. \ref{fig:HR} we show the evolution of our model in a Hertzsprung-Russell diagram from the zero-age main-sequence until the onset of the quiescent He-core burning phase. The evolution of convective zones, H- and He-burning luminosities and neutron densities are shown in Fig.  \ref{fig:kipp} from the He flash until the dredge up of the synthesized material to the surface. This time interval is highlighted in the evolutionary track of Fig. \ref{fig:HR}. After the He flash develops in the core, the outer boundary of the convective zone approaches the H-rich envelope. About 10 years after the He-flash, H starts to be engulfed in the He-flash driven convective zone. This feature of the evolution is usually called a proton ingestion episode (PIE). As a consequence of the engulfment of H into the hot He-burning zone, the reactions chain $^{12}{\rm C}(p,\gamma)^{13}{\rm N}(\beta)^{13}{\rm C}(\alpha,n)^{16}{\rm O}$ sets in and a flux of neutron is produced. The neutron density is shown color coded in Fig. \ref{fig:kipp}. 
About 180 yrs after the onset of the PIE, the neutron density reaches a maximum value of $\sim 5\times 10^{12}\,{\rm cm}^{-3}$, which is above the typical values for the $s$-process and on the low end of the range typical of the $i$-process nucleosynthesis. While $\sim\,$700 years after the onset of the PIE the convective zone splits into two due to the energy liberated by the H-burning, which makes the layers below it stable against convection. 
After its splitting, the convective zone is divided into two clearly separated regions, one deeper convective zone driven by He-burning and an outer convective zone driven by H-burning, with the consequent decoupling of the nucleosynthesis evolution of the two regions. In the inner convective zone, a neutron flux is still being produced until all the ingested protons are finally burned. As a consequence, $i$-process nucleosynthesis continues for some time in this region. However, nucleosynthesis products created here will not reach the surface. On the outer convective zone, H-burning continues after the splitting, yet temperatures here drop below the threshold for the $^{13}{\rm C}(\alpha,n)^{16}{\rm O}$ reaction to be efficient and the neutron density rapidly drops in this region. The production of trans-iron elements, and therefore their final photospheric abundances after 
dredge-up, are almost completely determined by the $i$-process nucleosynthesis until the moment of the splitting of the convective zone. The evolution of the PIE in our model is slow compared with other models of PIE in late He-core flashes, where the time span from the onset of the PIE until the splitting of the convective zone can be less than a year (see, e.g., Fig. 2 of \citealt{2008A&A...491..253M}). This time span depends upon the speed in which H is ingested into the convective zone. This, in turn, depends upon the intensity of the H-burning shell at the moment of the He flash, which itself is determined by the mass of the H-rich envelope at the moment of the He-flash. The lower the H-rich envelope mass, the faster the PIE. 
A thorough study of the dependence of nucleosynthesis on the H-mass at the moment of the He flash is being prepared for a follow-up paper. 
In the model presented here, about 200 years after the He-flash, a surface convective zone develops due to the evolution of the star to lower effective temperatures (see Fig. \ref{fig:HR}). This convective zone merges with the H-flash driven convective zone $\sim 4900$ years after the He flash, dredging up the processed material, rich in heavy elements, to the surface (see Fig. \ref{fig:kipp}). 

In Fig. \ref{fig:kiel} we show the evolution of the model during the He-subflashes that take place after the main He-core flash and the quiescent He-burning phase compared with the position in the Kiel diagram of the 11 known heavy-metal hot subdwarfs. The location of the quiescent He-burning phase of our simulation agrees well with the position of a subgroup of the stars. We note that a change in the initial composition of the model can shift the evolutionary track in both temperature and gravity. 

The surface abundance pattern that we obtain after the dredge up is shown in Fig. \ref{fig:abFe}, compared with the observed abundances of 5 of the heavy-metal hot subdwarfs for which there are abundances derived using NLTE atmospheric models \citep{2019A&A...630A.130D,2020A&A...643A..22D,2021A&A...653A.120D}. In this figure, we show the abundances with respect to iron ($[{\rm X}/{\rm Fe}]= \log(N_{\rm X}/N_{\rm Fe}) - \log(N_{\rm X}/N_{\rm Fe})|_{\odot}$, $N_{\rm X}$ being the abundance per number fraction of the element ${\rm X}$). 
It is particularly interesting that the abundance pattern of the iron-group elements that we obtain is  similar to those observed in the iHe-sdOB stars. The abundance pattern of the heavier trans-iron elements also follows relatively well the observations. The abundance of Pb of our model is higher than those measured in most iHe-sdOB stars, but agrees well with the abundance obtained for EC 22536-5304, the star in a binary system whose companion metallicity we have adopted for our model. However, the Pb abundance of this star in Fig. \ref{fig:abFe} might be underestimated as its Fe abundance is only an upper limit.

In Fig. \ref{fig:absun}, we present the abundances expressed as number fraction relative to solar values, $[{\rm X}/{\rm X}_{\odot}]= \log(N_{\rm X}/N_{\rm X,\odot})$, for all the elements observed in the iHe-sdOB stars. With the exception of Pb, the predicted abundances of heavy metals are generally lower than the observed values, although they follow a similar trend. Our model yields abundances for Sr, Y, Zr, and Sn that are approximately 5 to 10 times the solar abundance. These values are notably lower than the observed values in the iHe-sdOBs, but more than 2$\,$dex higher than the initial composition of the model, as depicted in Fig. \ref{fig:absun}. The initial composition of heavy elements are the solar abundances scaled to the metallicity of the model. The prediction of the model for Pb is significantly higher, approximately $10^4$ times solar. The high difference between the abundance of Pb and the abundance of other elements in our model is likely due to the specific choice of a low metallicity, where few seed nuclei are present to absorb neutrons, favoring the increase of Pb abundance at the expense of the abundance of lighter elements.

As our models do not account for element diffusion, the disagreement between our model abundances and those of the iHe-sdOB stars is not unexpected. The absence of non-equilibrium radiative levitation computations of trans-iron elements in the literature makes a direct comparison with the observations difficult. 
Yet, we will base our comparison on the few available diffusion models. 
Non-equilibrium diffusion models show that if diffusion acts unperturbed, a few hundreds of thousand years are enough to turn a purely He dominated surface into a hydrogen dominated one \citep{2008A&A...491..253M}. 
\cite{2011A&A...529A..60M} showed that to reproduce the He abundance of field sdOB stars additional mixing processes need to be included near the atmosphere. Under such conditions, He is depleted in the photosphere at timescales of the order of a few million years for the effective temperatures and gravities typical of the iHe-sdOB stars \citep{2011A&A...529A..60M}. This means that diffusion in these stars has been acting, at most, for a few million years. \cite{2011A&A...529A..60M} also demonstrated that for $T_{\rm eff}\simeq 30\,000$K elements heavier than Si (and lighter than Fe) will all be enhanced at the photosphere by the action of radiative levitation. These enhancements are slightly more than one order of magnitude for timescales of the order of a million years, with differences of a factor of a few depending on each particular element. If this effect were to be similar for trans-iron elements, it would imply that the abundances of elements heavier than Si ($Z=14$) would shift upwards by about an order of magnitude, while still preserving the upward trend displayed in Fig. \ref{fig:absun}. This, together with differences in the initial metallicities of the iHe-sdOB stars, would allow us to explain the observed abundances. 
The initial metallicity of our model reflects the metal-poor component of the Halo, whereas most Halo stars cluster around $[{\rm Fe}/{\rm H}]=-1.2$ \citep{2019ApJ...887..237C}, a value that could then be more representative of other iHe-sdOB stars. Also, the Halo-like kinematics of these stars could be explained if they are the surviving companions of type Ia SNe \citep{2021MNRAS.507.4603M}, in which case they could have higher initial metallicity. 
A higher initial metallicity in the models, if it does not suppress the PIE, could lead to higher abundances of trans-iron elements and a less pronounced peak in Pb. Further study is required to understand the production of trans-iron elements as a function of the initial composition of the star.

Our results indicate that heavy metals can be produced in a late He-flash model through $i$-process nucleosynthesis, which occurs as a consequence of hydrogen being ingested into the hot He-burning zone. This suggests that the high surface abundances of heavy elements in the iHe-sdOB stars may not solely be explained by radiative levitation. A self-synthesized origin for these observed abundances, possibly augmented by diffusion processes, appears plausible for the iHe-sdOB stars.

\section{Summary and conclusions}\label{sec:4}
In this letter we have calculated for the first time the detailed nucleosynthesis evolution of iron-group and trans-iron elements during the He-core flash of a red-giant star which has been stripped of its envelope. Specifically, we have calculated the evolution of a late He-flash model with an initial metallicity of $z=0.0004$ and an $\alpha$-enhanced composition, compatible with the composition of the companion star of EC 22536-5304. Our model undergoes $i$-process nucleosynthesis, producing heavy elements that are dredged up later to the surface, before the star settles into the quiescent He-burning phase. We have compared our results with the surface composition of heavy-metal iHe-sdOB stars. The abundance pattern of elements of the iron group and heavier elements of our model follows well the trend of the abundance pattern with atomic number observed in iHe-sdOB stars, but at lower absolute abundances. 
The latter can probably be linked to the absence of radiative levitation in our model, but also to the very low initial metallicity of our model. 
The exception is Pb, which is more abundant in our model than in the majority of the iHe-sdOBs. Our results indicate that neutron-capture processes can occur in a proton-ingestion episode following a He-flash in a model with a low-mass H-rich envelope, and that a significant amount of heavy elements, in particular lead, can be created. Such stellar structures might arise from a variety of binary and single evolutionary channels, future works will be aimed at characterizing them and their particular nucleosynthesis patterns. As mentioned in the introduction, one key question regarding the abundances of trans-iron elements observed in iHe-sdOB stars is whether enough material is present in neighboring layers for diffusion to act. Our computation shows that i-processes during a proton ingestion event offer a clear way to produce those trans-iron elements. Therefore, regardless of the action or not of radiative levitation, a self-synthesized origin of the heavy elements in the hot subdwarf stars is likely in the light of the present stellar evolution computations.

\begin{acknowledgements}
We thank the anonymous referee for their constructive criticism which helped to improve our work. T.B. thanks L. Siess for useful discussions about the performance of the \texttt{ANT} code and M. Dorsch for providing the abundances of the iHe-sdOB stars. 
M3B acknowledges grants PIP-2971 from CONICET (Argentina) and PICT 2020-03316 from Agencia I+D+i (Argentina). A.M.S. acknowledges grants PID2019-108709GB-I00 from Ministry of Science and Innovation (MICINN, Spain), Spanish program Unidad de Excelencia Mar\'{i}a de Maeztu CEX2020-001058-M, 2021-SGR-1526 (Generalitat de Catalunya), and support from ChETEC-INFRA (EU project no. 101008324). This research was supported by the Munich Institute for Astro-, Particle and BioPhysics (MIAPbP) which is funded by the Deutsche Forschungsgemeinschaft (DFG, German Research Foundation) under Germany's Excellence Strategy -- EXC-2094 -- 390783311.
\end{acknowledgements}

%
%

\bibliographystyle{aa} 
\bibliography{bibliografia}

\begin{appendix}
\section{Post-processing nucleosynthesis code: \texttt{ANT}}\label{appx:1}
\texttt{ANT} is a post-processing code for detailed nucleosynthesis computations. \texttt{ANT} reads at each stellar evolution time step the thermal structure and mixing velocities computed by a stellar evolution code, in the present paper these models are computed with \texttt{LPCODE}. Based on these input structures \texttt{ANT} then computes the detailed chemical evolution following nuclear burning and mixing with a very large number of species.  Convective mixing and burning is treated in a decoupled way in \texttt{ANT}. Between two \texttt{LPCODE} time steps, \texttt{ANT} can perform a number of sub time-steps alternating the computation of the nuclear reactions and the mixing of material. This approach should converge to a coupled treatment when the sub time steps converge to zero. 
In this work, we perform  with \texttt{ANT} 5 sub time steps between two \texttt{LPCODE} models. The time step of \texttt{LPCODE} in the proton ingestion episode is of the order of 15 minutes, and therefore, \texttt{ANT} time steps are of the order of 3 minutes. 

The nuclear network of \texttt{ANT} can comprise up to more than 5000 species. As stated above, for this work we have used a reaction network of 1190 nuclear species up to Po, including isotopes with a half-life time down to $0.5\,$ seconds. The equations for the nuclear reaction changes are solved using a Bader-Deufhard method \citep{bader1983semi}, which is basically a Bulirsch-Stoer method modified for stiff problems. The lineal algebra system is solved with the MA28 package \citep{reid1986direct}, that uses a direct method for sparse matrix. This implementation follows the suggestion of \cite{1999ApJS..124..241T}. The reaction rates for weak interactions (electronic captures and $\beta$ decays) are mainly taken from \cite{1987ADNDT..36..375T}, while all the other reaction rates are taken from the data base of JINA Reaclib \citep{2010ApJS..189..240C}. Neutron capture reactions follow the suggestion by the JINA Reaclib library, and corresponds to the KadOnis v.03 reaction rates when available \citep{dillmann2009kadonis,2006AIPC..819..123D} and theoretical reactions rates otherwise, mainly taken from \cite{2000ADNDT..75....1R,2010ApJS..189..240C,2010A&A...513A..61P}. In \texttt{ANT} we include the screening corrections for all the reactions involving two particles as reactants, following the works of \cite{1973ApJ...181..439D} and \cite{1973ApJ...181..457G}. The only reaction involving three particles for which we include an electron screening correction, is the $3\alpha$ reaction. We calculate this correction following the suggestion of \cite{1969ApJ...155..183S}. 

Convective treatment in \texttt{ANT} follows the scheme of \cite{2001ApJ...554.1159C}. Within this scheme, the mass fraction abundance of an isotope $k$ in a shell $i$ ($X^k_i$) is computed from the abundances before the mixing  of the same isotope $k$ in all the convective shells $j$ ($^0X^k_j$). After a time step $\Delta t$, the abundance $X^k_i$ is computed as
\begin{equation}
    X_i^k =\,^0\!X_i^k + \frac{1}{M_{\rm conv}} \sum_{j = {\rm conv}} (^0\!X_j^k-\,^0\!X_i^k)f_{ij} \Delta m_j,
\end{equation}
where $\Delta m_j$ is the mass of the shell $j$, and $M_{\rm conv}$ the total mass of the convective region. The factor $f_{ij}$ is defined as
\begin{equation}
    f_{ij} = \min\left(\frac{\Delta t}{\tau_{ij}},1\right),
\end{equation}
where $\Delta t$ is the time step in during which the material is mixed, and $\tau_{ij}$ is a characteristic time scale of convective motions between shells $i$ and $j$. The characteristic time $\tau_{ij}$ is defined as
\begin{equation}
    \tau_{ij} = \int_{r_i}^{r_j} \frac{dr}{v_{\rm MLT}}  = \sum_{l=i}^j \frac{\Delta r_l}{v_{l,{\rm MLT}}},
\end{equation}
where  $v_{l,{\rm MLT}}$ is the convective velocity at shell $l$, as provided by the mixing length theory. 

Initial abundances are taken from the first  \texttt{LPCODE} model read by \texttt{ANT} for the 32 species followed by \texttt{LPCODE} and completed for the other species with a solar scaled mixture at the corresponding metallicity.

\end{appendix}

\end{document}